# Modeling and Forecasting the Evolution of Preferences over Time: A Hidden Markov Model of Travel Behavior

May 11, 2017

**Feras El Zarwi** (corresponding author)
Department of Civil and Environmental Engineering
University of California at Berkeley
116 McLaughlin Hall
Berkeley, CA 94720-1720
feraselzarwi@gmail.com
+1 510 529 8357

**Akshay Vij**
Institute for Choice
University of South Australia
Level 13, 140 Arthur Street
North Sydney, NSW 2060
vij.akshay@gmail.com
+61 08 8302 0817

**Joan L. Walker**
Department of Civil and Environmental Engineering
University of California at Berkeley
111 McLaughlin Hall
Berkeley, CA 94720-1720
joanwalker@berkeley.edu
+1 510 642 6897

## Abstract

The motivation of this paper is to model and predict the evolution of preferences over time. Literature suggests that preferences, as denoted by taste parameters and consideration sets in the context of utility-maximizing behavior, may evolve over time in response to changes in demographic and situational variables, psychological, sociological and biological constructs, and available alternatives and their attributes. However, existing representations typically overlook the influence of past experiences on present preferences. This study develops, applies and tests a hidden Markov model with a discrete choice kernel to model and forecast the evolution of individual preferences and behaviors over long-range forecasting horizons. The hidden states denote different preferences i.e. modes considered in the choice set, and sensitivity to level-of-service attributes. The evolutionary path of those hidden states (preference states) is hypothesized to be a first-order Markov process such that an individual's preferences during a particular time period are dependent on their preferences during the previous time period. The framework is applied to study the evolution of travel mode preferences, or modality styles, over time, in response to a major change in the public transportation system. We use longitudinal travel diary from Santiago, Chile. The dataset consists of four one-week pseudo travel diaries collected before and after the introduction of Transantiago, a complete redesign of the public transportation system in the city. Our model identifies four modality styles in the population, labeled as follows: drivers, bus users, bus-metro users, and auto-metro users. The modality styles differ in terms of the travel modes that they consider and their sensitivity to level-of-service attributes (travel time, travel cost, etc.). At the population level, there are significant shifts in the distribution of individuals across modality styles before and after the change in the system, but the distribution is relatively stable in the periods after the change. In general, the proportion of drivers, auto-metro users, and bus-metro users has increased, and the proportion of bus users has decreased. At the individual level, habit formation is found to impact transition probabilities across all modality styles: individuals are more likely to stay in the same modality style over successive time periods than transition to a different modality style. Finally, a comparison between the proposed dynamic framework and comparable static frameworks reveals differences in aggregate forecasts for different policy scenarios, demonstrating the value of the proposed framework for both individual and population-level policy analysis.

# 1. Introduction

Discrete choice analysts have devoted much attention to the subject of preference heterogeneity. Preferences, as denoted by taste parameters and consideration sets in the context of utility-maximizing behavior, are regularly modeled as functions of demographic and situational variables. For example, value of travel time, or the marginal rate of substitution between travel time and cost in the context of travel mode choice, is frequently formulated as a function of income, and separate values of time are usually estimated for work and non-work travel (c.f. Parsons Brinkerhoff Quade & Douglas, Inc., 2005; Cambridge Systematics, 2002). Recent interest in the influence of latent psychological, sociological and biological constructs, such as attitudes, normative beliefs and affective desires, has led to the additional inclusion of these variables within existing representations of individual preferences (e.g. Bahamonde-Birke, 2015). Some studies have even contended that preferences are an endogenous function of the decision-making environment, as characterized by available alternatives and their attributes (e.g. Vij and Walker, 2014). Implicit to each of these representations is the following assumption: as these explanatory variables change over time, so should corresponding preferences.

However, most existing frameworks employ static representations of individual behavior that do not capture preference dependencies over time for the same individual. In addition to the variables identified previously, an individual's preferences in the present are expected to be a function of their preferences in the past, as evidenced by findings across multiple contexts, including transportation (Carrel et al., 2015), finance (Kaustia and Knüpfer, 2008), health (Gum et al., 2006), tourism (Sönmez and Graefe, 1998), sustainable development (O'Hara and Stagl, 2002), etc. Notwithstanding this evidence, discrete choice frameworks that capture such temporal dependencies are rare in the literature. Part of the limitation is empirical: most studies use cross-sectional data, and longitudinal data of the kind that is needed is not always available.

The ability to understand and predict how individual preferences evolve over time offers the potential to address transportation policy questions of great interest. Who is more likely to use shared mobility services: individuals who currently drive, or those who take public transport? Will the adoption of driverless cars be led by individuals with significant past exposure to other new technologies, or individuals with the greatest need for access to self-driving car technology? How do changes to the public transport system impact individuals that are differently predisposed towards available travel modes? Transport system use and policy will vary, often considerably, depending upon the answer to each of these questions. In fact, it is this last question that motivates the empirical application in this study.

The objective of this study is to develop an econometric framework that can model preference dependencies over time for the same individual. Our proposed framework constitutes a hidden Markov model (HMM) with a discrete choice kernel. Decision-makers are assumed to be utility-maximizing, and the unobserved states denote different preferences, as denoted by differences in taste parameters and consideration sets. Transitions between preferences are expressed as a function of time-varying covariates, namely socio-demographic variables and alternative attributes. The evolutionary path is hypothesized to be a first-order Markov process such that an individual's preferences during a particular time period are dependent on their preferences during the previous time period. The framework is empirically evaluated using data from the Santiago Panel (Yáñez et al., 2010), which comprises four one-week waves of pseudo travel-diary data spanning a twenty-two month period that extends both before and after the introduction of Transantiago, a major redesign of the public transport system in Santiago, Chile.

HMMs were first proposed nearly five decades ago (Baum et al., 1970; Baum and Petrie, 1966). They have a rich history of application in machine learning, with particular regards to the subject of speech recognition (Rabiner, 1989). They have also been applied, albeit limitedly, to the study of individual behavior in the applied disciplines of education (e.g. Hong and Ho, 2005; George, 2000), marketing (e.g. Netzer et al., 2008) and transportation (e.g. Xiong et al., 2015; Choudhury et al., 2010; Goulias, 1999). Our contribution in this paper is to develop, apply, and test an HMM framework that captures, models and forecasts the evolution of individual preferences and behaviors over long-range forecasting horizons.

The remainder of the paper is organized as follows: Section 2 motivates the study through a discussion of previous findings on the evolution of individual preferences over time; Section 3 reviews dynamic discrete choice model frameworks that have been used in the past to model temporal interdependencies in preferences and behavior, and how they relate to our proposed HMM framework; Section 4 outlines the proposed methodological framework; Section 5 discusses the initial conditions problem in dynamic discrete choice models, and if and how it applies to HMMs; Section 6 describes the dataset that constitutes our empirical application; Section 7 presents results from the model framework; Section 8 demonstrates the benefits of the framework for policy analysis; and finally, Section 9 concludes with a discussion of key findings, limitations and directions for future research.

## 2. Motivation: Evolution of Individual Preferences over Time

Most economists would agree that individual preferences, as denoted by taste parameters and consideration sets in the context of utility-maximizing behavior, can and do change over time. However, most would also contend that understanding why particular preferences exist in the first place, and consequently, how they change over time, ought not to be the concern of mainstream economics. While the view has been challenged over the years (notable examples include Becker, 1996 and Elster, 2016), most contemporary economic representations of individual behavior continue to treat preferences as exogenously determined, and attention is usually limited to understanding and predicting policy implications under any given set of preferences.

Preferences may change over time in response to changes in, among others, demographic and situational variables, psychological, sociological and biological constructs, and available alternatives and their attributes. Changes in preferences have been observed across a broad spectrum of behavioral contexts, from the personal to the public. For example, Buss et al. (2001) examined the evolution of mate preferences between 1939 and 1996 at geographically different locations in the United States. Their findings indicate that mate preferences did indeed change. When looking for a potential partner over time, both males and females increased the importance of physical attraction and financial status, and males decreased the importance of domestic skills. At the other end of the spectrum, Page and Shapiro (1982) studied the evolution of preferences on matters of domestic and foreign policy, such as civil liberties, abortion, etc., between 1935 and 1979 in the United States. They found that significant shifts in preferences were rarely the case over short time periods. However, when opinions and preferences did actually change, that was the outcome of changes occurring in the decision-making environment, whether in the social and economic spectrum or in the lives of decision-makers.

In the context of transportation, perhaps the 'peak car' phenomenon best represents the notion of changing preferences over time. The turn of the twenty-first century has witnessed stagnant or declining levels of car use across much of the developed world (Goodwin and Dender, 2013; Garceau et al., 2014). The shift in preferences away from the car as a mode of transportation has been attributed to a combination of economic, social and technological factors that include a recessionary global economy, fluctuating oil prices, ageing national populations, shifts in cultural values, advances in information and communications technology, etc. (see, for example, Vij et al., 2017; McDonald, 2015; Kuhnimhof et al., 2013; Collet, 2012).

What about travel behavior in the era of transformative mobility? Why would one expect preferences to change over time in response to major changes in the transportation system, such as the introduction of autonomous vehicles? There may be changes in consideration sets. Individuals unwilling or unable to drive themselves may be willing and able to use autonomous vehicles. There may be changes in taste parameters. Being in an autonomous vehicle will allow decision-makers to multitask, which may cause them to be: (1) less sensitive to driving during peak hours and getting caught up in congestion; (2) not worried about finding a parking spot in congested cities nor paying parking fees; and (3) more flexible in terms of residential choice location as they might consider residing outside dense urban cities and commute via the autonomous vehicle since driving has become less onerous. These factors may lead to changes in value of time (VOT). The assumption that preferences are stable may be valid when forecasting over short-term periods. However, when forecasting over long-term horizons, we need to take into account that various shocks/changes in the built environment and investments in technologies and services are bound to happen, and that these shocks/changes will likely impact preferences.

Preferences may additionally depend upon past experiences. Though most neoclassical frameworks assume that preferences are inter-temporally separable, studies on the formation and persistence of habits have questioned the validity of the assumption (Muellbauer, 1988; von Weizsäcker, 1971; Pollak, 1970). Past experiences provide a ready yardstick for comparison, serving both to magnify differences under certain contexts, and reduce contrasts in others. As Becker (1992) writes, "a given standard of living usually provides less utility to persons who had grown accustomed to a higher standard in the past. It is the decline in health, rather than simply poor health, that often makes elderly persons depressed. And what appeared to be a wonderful view from a newly occupied house may become boring and trite after living there for several years."

Past experiences can also serve as anchors, dampening the ability of external events to force commensurate shifts in individual preferences. Two individuals with completely exchangeable current circumstances may still differ in terms of their preferences, due to corresponding differences in their personal histories and the life paths that brought them here. For example, Bronnenberg et al. (2012), in their study on the long-run evolution of brand preferences among individual consumers, concluded that "brand capital evolves endogenously as a function of consumers' life histories and decays slowly once formed". Their findings are echoed by studies in other behavioral contexts. Travel behavior in particular, due to its repetitive nature, is especially prone to habit formation (Thøgersen, 2006; Gärling and Axhausen, 2003; Sönmez and Graefe, 1998; Aarts et al., 1997). "Habits, once formed, become regularized and the market mechanism virtually ceases to operate", and "consequently, if these habits can be identified, choices made at any future decision point can be predicted with a fairly high degree of accuracy" (Banister, 1978). As an extreme example, some studies have speculated that the use of active modes of transportation (i.e. walking and bicycling) as children can promote more sustainable travel behavior practices as adults (see, for example, Mitra et al., 2010; Faulkner et al., 2009; Roberts, 1996).

However, hypotheses such as these have rarely been tested in the literature, due largely to limitations on available data. Transportation planning has typically relied on cross-sectional mobility data for understanding and predicting different dimensions of travel and activity behavior. Cross-sectional studies can provide population snapshots at a point in time; by extension, repeated cross-sections can show broad population trends over time. However, cross-sectional studies cannot measure changes at the level of the individual over time. As mentioned before, the ability to understand and predict changes in individual-level preferences and behaviors offers the potential to address transportation policy questions of great interest.

Consider, for example, the peak car phenomenon. A 5% decrease in driving mode shares at the population level over time could imply that 5% of the population has stopped driving, or that the entire population is driving 5% less, or some combination of the two (Hanson and Huff, 1988). The nature and impact of transport policy will depend on which of these competing hypotheses is true; unfortunately, a traditional cross-sectional study would be unable to distinguish between these hypotheses. Similarly, consider the case of new transportation technologies and services, such as autonomous and/or alternative-fuel vehicles and shared mobility services, that promise to transform mobility. The diffusion of new technologies and services is a temporal process (Rogers, 2010). The key to understanding the future of mobility is not only to study the immediate impact of current policies, services, and nudges; but also how these impacts influence trends and their evolution over decades, particularly as new technologies and services are introduced. For example, is the growth in carsharing and ridesharing services being led by individuals who have always been multimodal, or do these services also appeal to car-dependent households? Will self-driving cars be subject to the constraints of an ownership-based economy, or will gradual changes in preferences imply that access and use is facilitated primarily through shared services? Cross-sectional studies that use static frameworks cannot address these questions. Where such insight is required, longitudinal studies that use dynamic frameworks are necessary.

## 3. Methodological Basis: Dynamic Models for Discrete Choice Analysis

Dynamic discrete choice models try to account for the influence of past experiences on present choices. According to Kenneth Train (2009), current choices affect future choices, as past choices affect current choices, and this causality provides the basis for dynamic discrete choice modeling. There are two broad paradigms in the literature (for an excellent synthesis on the subject, the reader is referred to von Auer, 1998). Both paradigms assume that present preferences and behavior are impacted by past experiences; they differ in the ascribed importance of expected future utility on present behavior.

The first paradigm assumes that individuals, when making a decision at a given time period, behave as if they are forward-looking agents that maximize their present and expected future discounted utility over the entire time horizon. Perhaps the most famous example of such a representation of dynamic discrete choice behavior is the study by Rust (1987) on the optimal replacement of bus engines. Rust's representation has since been applied to many contexts, including car ownership (see for example Cirillo and Xu, 2011; and Glerum et al., 2013), and it is in this context that we describe the framework. A car is considered a durable good that yields utility over time. An individual's choice of whether to purchase a car at a certain time period or postpone the purchase depends on how that individual expects to use the car both now and in the future.

The second paradigm assumes a more myopic view of behavior, where individuals are assumed to maximize their present utility, and future expected utility is completely discounted. In other words, the individual cares only about the current time period, and choices in later time periods are deemed irrelevant. For theoretical treatments of such myopic representations of individual behavior, the reader is referred to, among others, Gorman (1967), Pollak (1970) and von Weizsäcker (1971). The HMM conforms to this second paradigm, where an individual's preferences in the present are assumed to be dependent on their preferences in the past, but at any given point in time, the individual is assumed only to maximize present utility.

Depending on the empirical context, one or the other paradigm may be preferred. When studying medium and long-term travel and activity behaviors, such as car ownership and residential location, it may be more reasonable to assume that individuals are forward-looking. Decisions such as whether to buy a car and where to live have implications that extend well beyond the present. However, when studying short-term travel and activity behaviors, such as travel mode choice, it may be more reasonable to assume that individuals are myopic. The impact of these decisions is typically short-lived and readily reversible. Since our model framework will be applied to the study of short-term behaviors, we will be adopting a myopic view of decision-making, articulated through the HMM framework.

As mentioned before, HMMs have been used previously to study the dynamics of travel and activity behavior. Goulias (1999) used HMMs to study the dynamics of household time allocation where the dependent variable is continuous. Choudhury et al. (2010) used HMMs to represent the evolution of latent plans over time, and their consequent impact on actions at any particular point in time. Their framework does make an explicit link with discrete choice analysis. They apply their framework to model the "evolution of unobserved driving decisions as drivers enter a freeway." Their model is described very generally; extensions such as incorporating the expected maximum utility are not implemented and applications to long-range modeling and forecasting are not investigated. Perhaps the empirical application that is closest to the work presented here is the study by Xiong et al. (2015), who used HMMs to study the dynamic nature of travel mode choice behavior over time. Their framework does not allow for heterogeneity with regards to consideration sets, the transition model is not sensitive to changes in available alternatives and their attributes, and the value of the framework for policy analysis, beyond improvements in fit, is unclear. Our objective is to build upon these previous studies to develop a methodological framework capable of modeling the dynamics of preferences over time in a manner that is theoretically grounded, behaviorally meaningful and practically useful.

## 4. Methodological Framework

Our methodological framework builds on dynamic models, which are becoming more popular in the field of travel behavior. For example, Van Acker et al. (2014) highlight the need for incorporating dynamics into models of behavior by stating that "it will almost inevitably be the case that the range of travel choices open to people will be wider over time periods in which lifestyles can also change, than in the short run when the constraints will be more prominent. As such, the whole way of thinking about travel and lifestyle must be seen as a process of change over time, not as a fixed state".

We propose using a hidden Markov model (HMM) with a discrete choice kernel, where the following two key assumptions are made: (1) we assume a myopic view of behavior, such that observed choices at a certain time period t are only dependent on corresponding preferences during that time period, and future expected utility is completely discounted; and (2) the hidden states denote different preferences, and the evolution of preferences over time is assumed to be a first-order Markov process such that an individual's preferences during a certain time period is dependent on their preferences during the previous time period. Figure 1 illustrates the HMM assumptions. It is important to note that 'preference state at time 1' determines the effects of inertia and past experiences on the probabilistic assignment of each individual to a particular set of preferences during the first time period.

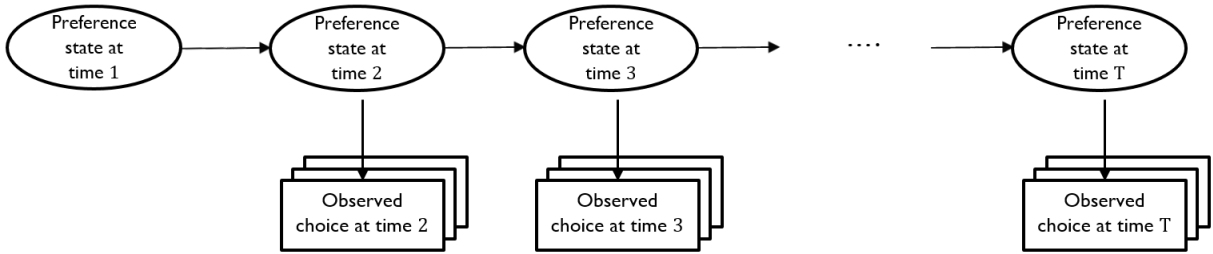

**Figure 1: Hidden Markov Model Structure (figure adapted from Choudhury et al., 2010)**

Hidden Markov models comprise three components: initialization model, transition model, and observed output model (Jordan, 2003). The initialization model predicts the probability that a decision-maker belongs to a certain hidden state during the first time period. The transition model predicts the probability of observing a certain evolution of hidden states between successive time periods. Lastly, the observed output model predicts the probability of observing a vector of choices for a decision-maker at a given time period, conditional on belonging to a certain hidden state during that time period.

We operationalize the HMM in the context of travel mode choice behavior by relying on the construct of modality styles. The construct has been introduced in the literature to refer to overarching lifestyles, built around the use of a particular set of travel modes, that influence all dimensions of an individual's travel and activity behavior (Vij et al., 2013). In the context of travel mode choice behavior, we use modality styles to refer to distinct segments in the population with different travel mode preferences, i.e. modes considered in the choice set, and sensitivity to level-of-service attributes. For example, modality style models have shown that in 2000, 42% of the San Francisco Bay Area's population exclusively considered driving, whereas this share reduced to 23% in 2012 (Vij et al., 2017). Investment in technologies and services are expected to influence both the travel modes considered and the sensitivity to level-of-service attributes. Consider, for example, the case of autonomous vehicles. A fully autonomous vehicle that is capable of navigating itself without human input might prompt changes in the value of time, through its ability to allow passengers to engage in whatever tasks they wish to while inside the car. Similar changes in preferences can be imagined in response to other changes in the transportation system. Modeling what types of modality styles have flourished or declined over time is key to understanding and predicting mode share shifts in response to policies, services, technologies and nudges.

Accordingly, in the context of travel mode choice behavior, the unobserved states in the dynamic framework shall be represented by modality styles. Through the remainder of the paper, we will use the terms modality styles and (travel mode) preferences interchangeably. The transition model quantifies the evolution of modality styles over time to capture structural shifts in preferences. Our dynamic framework requires a transition model that can capture shifts in modality styles brought about by major changes to the transportation system (sharing, automation, transit on demand) or by shifts in attitudes (e.g. towards/away from auto-orientation), or changes in socio-demographic variables. We are interested in forecasting, and thus require a structural model for the transition probabilities that captures the influence of transportation and societal changes. For this we will employ a homogenous HMM, which assumes that the transition model between modality styles (preferences) from one time period to the other is consistent/static i.e. the parameters entering the transition model between subsequent waves are time-invariant. Any differences in transition probabilities over waves are assumed to arise due to changes in the explanatory variables entering the transition model. Figure 2 displays the dynamic nature of our framework. Over following subsections, we explain each of the constituent sub-models in greater detail.

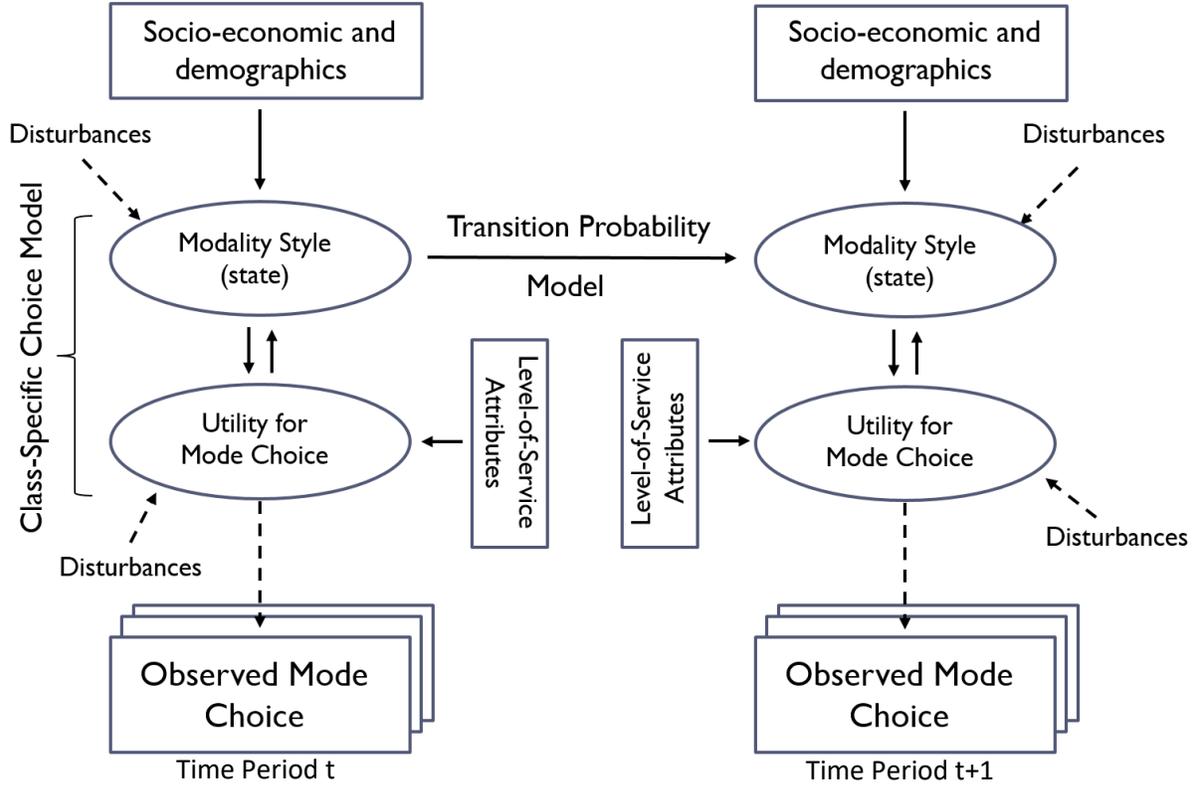

**Figure 2: Proposed Dynamic Discrete Choice Framework**

### 4.1 Class-specific Mode Choice Model

The class-specific mode choice model predicts the probability that individual *n* during time period *t* made a set of choices $y_{nt}$, conditional on the individual belonging to modality style, or class, *s* during that time period. Note that $y_{nt}$ is a vector whose element $y_{ntkj}$ equals one if the individual chose travel mode *j* during choice situation *k* over time period *t*, and zero otherwise. The model allows more than one choice situation per individual and time period, and correlation between these choice situations is captured through the assumption that an individual's modality style remains stable over a single time period.

Let $U_{ntkj|s}$ denote the utility of travel mode *j* during choice situation *k* over time period *t* for individual *n*, conditional on the individual belonging to modality style *s*, and is expressed as follows:

$$U_{ntkj|s} = V_{ntkj|s} + \varepsilon_{ntkj|s} = x'_{ntkj}\beta_s + \varepsilon_{ntkj|s}$$

where $V_{ntkj|s}$ is the systematic utility, $x'_{ntkj}$ is a row vector of attributes of alternative *j* during choice situation *k* over time period *t* for individual *n*, $\beta_s$ is a column vector of parameters specific to modality style *s* and $\varepsilon_{ntkj|s}$ is the stochastic component of the utility specification. Now, assuming that all individuals are utility-maximizers and $\varepsilon_{ntkj|s}$ follows an i.i.d. Extreme Value Type I distribution across individuals, time periods, choice situations, alternatives and modality styles with location zero and scale one, the probability that individual *n* chooses travel mode *j* during choice situation *k* over time period *t*, conditional on the individual belonging to modality style *s*, is as follows:

$$P(y_{ntkj} = 1|q_{nts} = 1) = P(U_{ntkj|s} \geq U_{ntkj'|s} \, \forall \, j' \in C_{ntk|s}) = \frac{e^{x'_{ntkj}\beta_s}}{\sum_{j' \in C_{ntk|s}} e^{x'_{ntkj'}\beta_s}}$$

where $P(y_{ntkj} = 1|q_{nts} = 1)$ denotes predicting the probability that individual $n$ over wave $t$ and choice situation $k$ chooses alternative $j$ (implying $y_{ntkj}$ equals one and zero otherwise) conditional on belonging to modality style s during wave t ($q_{nts}$ equals one and zero otherwise), and $C_{ntk|s}$ denotes the choice set available for individual $n$ at wave $t$ and choice situation $k$ conditional on modality style $s$. Preference heterogeneity is captured by allowing both the taste parameters $\beta_s$ and the consideration sets $C_{ntk|s}$ to vary across modality styles.

Assuming that choice probabilities for individual $n$ across all choice situations belonging to time period $t$ are conditionally independent, given that the individual belongs to modality style $s$ during time period $t$, the conditional probability of observing a vector of choices $y_{nt}$ for a certain time period $t$ becomes:

$$P(y_{nt}|q_{nts} = 1) = \prod_{k=1}^{K_{nt}} \prod_{j \in C_{ntk|s}} P(y_{ntkj} = 1|q_{nts} = 1)^{y_{ntkj}}$$

where $K_{nt}$ is the number of distinct choice situations observed for individual $n$ over time period $t$.

### 4.2 Initialization Model
The initialization model predicts the probability that individual $n$ belongs to modality style $s$ during the first time period. The probabilities are expressed as a function of individual characteristics during that time period, denoted by the column vector $z_{n1}$. Characteristics may include observable socio-economic and demographic variables, such as income and gender, or later psychological, sociological or biological constructs, such as attitudes, normative beliefs or affective desires. In our case, information on latent constructs was not available across all observation periods, and characteristics include observable socio-economic and demographic variables only. Depending on the analyst's assumption, the model may be formulated as a multinomial logit, multinomial probit, mixed logit or some other model form. We assume that the initialization model is multinomial logit.

Let $U_{n1s}$ denote the utility of modality style $s$ during the first wave for individual $n$ which is expressed as follows:

$$U_{n1s} = V_{n1s} + \varepsilon_{n1s} = z'_{n1}\tau_s + \varepsilon_{n1s}$$

where $V_{n1s}$ is the systematic utility that is observed by the analyst, $z'_{n1}$ is a row vector of socio-economic and demographic variables for individual $n$ during the first wave and $\tau_s$ is the associated column vector of parameter estimates for modality style $s$, and $\varepsilon_{n1s}$ is the stochastic component of the utility specification. Now, assuming that all individuals are utility maximizers and that $\varepsilon_{n1s}$ follows an i.i.d. Extreme Value Type I distribution across individuals, first wave, and modality styles with location zero and scale one, the initialization model could be formulated as such:

$$P(q_{n1s} = 1|Z_{n1}) = P(U_{n1s} \geq U_{n1s'} \, \forall \, s' = 1,2,\ldots,S) = \frac{e^{z'_{n1}\tau_s}}{\sum_{s'=1}^{S} e^{z'_{n1}\tau_{s'}}}$$

where $P(q_{n1s} = 1|Z_{n1})$ represents the probability that individual *n* has modality style *s* during the first wave conditional on his/her socio-demographic variables during the first wave, and $S$ denotes the total number of modality styles in the sample.

**4.3 Transition Model**

Analogously, the transition model predicts the probability that individual *n* transitions to modality style *s* during time period *t*, conditional on the individual belonging to modality style *r* during the previous time period *(t-1)*. Ordinarily, the probabilities may be expressed as a function only of individual characteristics during that time period (see, for example, Xiong et al., 2015), as was the case with the initialization model. Depending on the analyst's assumption, the transition model may be formulated as a multinomial logit, multinomial probit, mixed logit or some other model form. We assume that the transition model is multinomial logit.

Let $U_{nts|(t-1)r}$ denote the utility derived from transitioning into modality style *s* at wave *t* conditional on individual *n* belonging to modality style *r* during the previous wave *(t-1)*, which is expressed as follows:

$$U_{nts|(t-1)r} = V_{nts|(t-1)r} + \varepsilon_{nts|(t-1)r} = z'_{nt}\gamma_{sr} + \varepsilon_{nts|(t-1)r}$$

where $V_{nts|(t-1)r}$ is the systematic utility, $z'_{nt}$ is a row vector of observable socio-economic and demographic characteristics of individual *n* over wave *t* and $\gamma_{sr}$ is a column vector of parameters specific to modality style *s* at wave *t* given that the individual belonged to modality style *r* during wave *(t-1)*, and $\varepsilon_{nts|(t-1)r}$ is the stochastic component of the utility specification.

Assuming that all individuals are utility maximizers and that $\varepsilon_{nts|(t-1)r}$ follows an i.i.d. Extreme Value Type I distribution across individuals, waves and modality styles with location zero and scale one, the transition probability could be formulated as such:

$$P(q_{nts} = 1|q_{n(t-1)r} = 1) = P(U_{nts|(t-1)r} \geq U_{nts'|(t-1)r} \, \forall \, s' = 1,2,\ldots,S) = \frac{e^{z'_{nt}\gamma_{sr}}}{\sum_{s'=1}^{S} e^{z'_{nt}\gamma_{s'r}}}$$

where $P(q_{nts} = 1|q_{n(t-1)r} = 1)$ denotes one entry of the transition probability matrix, which involves predicting the probability that individual *n* belongs to modality style *s* during wave *t*, for *t > 1*, conditional on modality style *r* during the previous wave *(t-1)*.

Now, the transition model is merely a function of socio-demographic variables. However, wouldn't changes in the level-of-service of the transport network, such as reductions in travel times or travel costs, influence the transition from one modality style to the other? Changes in the level-of-service of different travel modes will affect different modality styles differently. For example, increased freeway congestion will make car-oriented modality styles less attractive, and a reduction in transit services will have a similar effect on transit-oriented modality styles. These changes will likely impact whether and how individuals change their modality styles, and should be accordingly captured by the transition model. We account for these changes by formulating transition probabilities as an additional function of the consumer surplus each individual would derive by belonging to different modality styles (building off the static framework forwarded by Vij and Walker, 2014).

Given that individuals are assumed to be utility-maximizing, the consumer surplus offered by modality style *s* to individual *n* during time period *t* is given theoretically by the total expected maximum utility derived by the individual over all observations for that time period, also referred to as the inclusive value.

When the class-specific choice model is assumed to be multinomial logit, expected maximum utility reduces to the familiar logsum measure, and the average consumer surplus is given by:

$$CS_{nts} = \frac{1}{K_{nt}} \sum_{k=1}^{K_{nt}} \log \left[ \sum_{j \in C_{ntk|s}} e^{x'_{ntkj}\beta_s} \right]$$

The transition probability we are proposing is defined as follows:

$$P(q_{nts} = 1 | q_{n(t-1)r} = 1) = \frac{e^{z'_{nt}\gamma_{sr} + CS_{nt}\alpha_{sr}}}{\sum_{s'=1}^{S} e^{z'_{nt}\gamma_{s'r} + CS_{nt}\alpha_{s'r}}}$$

where $\alpha_{sr}$ is a parameter associated with the consumer surplus specific to modality style *s* at wave *t* given that the individual belongs to modality style *r* over wave (*t-1*). For the model to be consistent with utility-maximizing behavior, $\alpha_{sr} \geq 0$.

The inclusion of consumer surplus in the transition model provides a basis for understanding and predicting how individual preferences might change over time in response to corresponding changes in the transportation system. Consider, for the sake of illustration, that the local public transport agency introduces a temporary free pass for all services. The introduction of such a pass would change the consumer surplus offered by different modality styles differently. For modality styles that do not include public transport in their consideration set, the consumer surplus will be unchanged. For modality styles that do include public transport, consumer surplus will be higher, making individuals more likely to belong to these modality styles in the subsequent time period. In particular, the greatest change will likely be for a modality style that both considers public transport and is highly sensitive to travel costs (since the free pass will impact travel costs). Therefore, the introduction of the free pass might not only lead individuals to expand their consideration sets, it may cause them to become more sensitive to travel costs. Similar changes could potentially be modeled for other scenarios. This is a key benefit to our framework.

### 4.4 Likelihood Function of the Full Model

Now, the marginal probability $P(y_n)$ of observing a sequence of choices $y_n$ for decision-maker *n* over *T* time periods is expressed as follows:

$$P(y_n) = \sum_{s_1=1}^{S} \sum_{s_2=1}^{S} \cdots \sum_{s_T=1}^{S} \prod_{t=1}^{T} P(y_{nt}|q_{nts_t} = 1) P(q_{n1s_1} = 1|Z_{n1}) \prod_{t=2}^{T} P(q_{nts_t} = 1|q_{n(t-1)s_{t-1}} = 1)$$

HMMs are traditionally estimated via the Expectation-Maximization (EM) algorithm (forward-backward algorithm) that provides a computationally robust method of optimization by taking advantage of the conditional independence properties of the model framework. The EM algorithm is particularly useful for HMMs because in the M-step, each of the class-specific choice models, the initialization model and transition model can be maximized independently. However, for HMMs that incorporate feedback to the transition model through the construct of consumer surplus, this will no longer be the case. The class-specific choice model and the transition model can no longer be maximized independently in this case, since the class-specific taste parameters are common to both sub-models. Consequently, the EM algorithm is not useful in this case, and we resorted to using traditional batch gradient optimization techniques.

## 5. Initial Conditions Problem in Dynamic Discrete Choice Models

Dynamic models may exhibit what is referred to as the initial conditions problem, first discussed by Heckman (1981). The initial conditions problem refers to how the dynamic process is initialized. Heckman (1981) illustrates the problem with the following functional form:

$$U_{nt} = f(x_{nt}, y_{n1}, y_{n2} \dots y_{n(t-1)}) + \varepsilon_{nt}$$

where $U_{nt}$ denotes the utility during time period *t* for individual *n*, $f$ denotes the function that expresses the observable components of utility, $x_{nt}$ entails explanatory variables for time period *t* for individual *n*, $y_{n(t-1)}$ represents the choice that individual *n* made during time period *(t-1)*, and $\varepsilon_{nt}$ is the stochastic error component of the utility specification. We can clearly see that this general dynamic model captures the effect of previous choices on current ones.

One main assumption in this model formulation entails serially correlated error structure. According to Heckman (1981), initial conditions can only be treated as exogenous variables if at least one of the following two conditions is met: (1) serially independent error structures in the model framework ($\varepsilon_{nt}$) whereby the error components are assumed to be independently and identically distributed over time; or (2) if the data includes observations since the dynamic process started. If one of these two conditions is met, then we can treat initial conditions as exogenous variables or "fixed". However, if neither of those assumptions is met, then initial conditions cannot be treated as exogenous variables, and assuming that they are will lead to inconsistent parameter estimates. The latter condition is almost never going to be met, since the analyst frequently only observes a dynamic process after it first began. Therefore, in our case, in order to treat the initial conditions as exogenous, the first condition must hold true.

Heckman (1981) discusses this issue in the context of a fixed effect probit model. Let us reframe our proposed hidden Markov model using the notation employed by Heckman's general dynamic model. The initial conditions problem, in the case of HMMs, is associated with the initialization model with left-censored datasets. The class-specific choice model could be expressed as follows:

$$U_{nt} = f(x_{nt}, q_{nt}) + \varepsilon_{nt}$$

where $x_{nt}$ entails explanatory variables for time period *t* for individual *n*, $q_{nt}$ denotes the modality style for decision-maker *n* during time period *t*. However, according to the aforementioned transition model, one could express the following:

$$q_{nt} = g(z_{nt}, q_{n(t-1)}) + \tau_{nt}$$

where $g$ denotes the function that models the transition of modality styles over time, $z_{nt}$ entails socio-demographic variables for time period *t* for individual *n*, and $\tau_{nt}$ is the stochastic error component. Accordingly, we can express the class-specific choice model utility equation as such:

$$U_{nt} = f'(x_{nt}, z_{nt}, q_{n(t-1)}) + \varepsilon_{nt}$$

We can recursively iterate by replacing the expression of modality styles at different time periods all the way until the first time period, ending up with the following equation:

$$U_{nt} = f''(x_{nt}, z_{n1} \dots z_{nt}) + \varepsilon_{nt}$$

We are assuming that the choice probabilities that comprise the class-specific choice model for a certain individual are conditionally independent over choice situations and time periods, given the modality style they belong to and the set of explanatory variables that affect the choice process. Thus, the error components in this choice model are independently and identically distributed across time, i.e. no serial correlation. Therefore, by assuming serially independent error structures, which is standard in HMMs, initial conditions could be treated as exogenous variables or "fixed" for the aforementioned reasons.

Conditional on an individual's modality style, how valid is it to assume that the utilities of different choice alternatives over time are serially uncorrelated? Factors that lead to serially correlated error terms entail habit or inertia whereby choices (travel patterns in the case of travel behavior) could repeat themselves over time (Cantillo et al., 2007; Gärling and Axhausen 2003). The construct of modality styles tries to capture profound individual variations in preferences and attitudes and "higher-level orientations, or lifestyles that influence all dimensions of an individual's travel and activity behavior" (Vij, 2013). We assume that by conditioning on those higher-level lifestyle orientations, or modality styles, we are fully accounting for habit or inertia effects.

Another factor behind serial correlation in the case of panel data comprises multiple choice decisions made by the same individual. This is what we refer to as specification bias, which encompasses excluding important determinants of choice decisions that are unobserved by the analyst but are common across multiple choice decisions for the same individual. These determinants could include unobserved attitudes, missing socio-demographic variables, etc., which become confounded with the error terms over time and could induce the main source of correlation between choice decisions made by the same individual over time. Our hypothesis is that this shared correlation is captured through the construct of modality styles. That is why, once we control for those higher-level orientations, it becomes reasonable to assume that choices are serially independent.

For a hidden Markov model, the evolutionary path, i.e. the transition model, is depicted by a first-order Markov process, which follows the property:

$$\pi^n = \pi^1 \prod_{t=2}^{T} \Omega_{t-1,t}$$

where $\pi^n$ denotes the vector of marginal probabilities for the available modality styles at time period n, $\pi^1$ has the same definition as $\pi^n$ but is associated with the first time period, and $\Omega_{t-1,t}$ denotes the transition probability matrix between time period (*t-1*) and *t*.

If the data is left-censored whereby the time periods which were observed correspond to: {J, J+1, ..., T}, such that J>1, then the initialization model will be biased. Using the above Markov chain equation, the initialization probabilities evaluated at t=J equal the product of the initialization probabilities at t=1 and all the transition probabilities up until t=J. In other words, $\pi^J = \pi^1 \times \prod_{t=2}^{J} \Omega_{t-1,t}$, and the magnitude of the bias is given by the difference between $\pi^J$ and $\pi^1$. However, the transition model and the class-specific choice models will remain unbiased. Our main objective in this research paper is to develop a framework for modeling and forecasting the evolutionary path of preferences over time. In order to do so, it is important that parameter estimates associated with the transition model and class-specific choice model remain unbiased. Therefore, the initial conditions problem that exists in other types of dynamic models is not of concern here.

We conducted a Monte Carlo simulation experiment to corroborate our arguments. For our Monte Carlo simulation, we simulated 5000 observations. Each of the observations entailed 10 time periods. There were two available states ($s_1, s_2$) that each observation could belong to at each time period. There were also two available outcomes at each time period. The distribution of the initialization model, during the first time period, is 0.4 and 0.6 across the two states i.e. $\pi^1 = [0.4, 0.6]$. The transition probability matrix between two successive time periods is $\Omega_{t-1,t} = \begin{bmatrix} 0.8 & 0.2 \\ 0.3 & 0.7 \end{bmatrix}$. The class-specific choice probabilities were assumed as follows: conditional on being in the first state, the probability of choosing the first outcome is 0.5, and the probability of choosing the second outcome is 0.5. However, conditional on being in the second state, the probability of choosing the first outcome is 0.7, and the probability of choosing the second outcome is 0.3. We first estimated the hidden Markov model parameters for N=5000 observations over the entire time periods associated with the dynamic process i.e. T=10. We then re-estimated the HMM parameters by truncating the dataset by removing the first 5 time periods for each of the 5000 observations. Table 1 summarizes the results from the model estimation using the Expectation – Maximization algorithm. The initialization model for N=5000 and T=5 could be computed as such: $\pi^6 = \pi^1 \times \prod_{t=2}^{6} \Omega_{t-1,t} = [0.59, 0.41]$. We can clearly observe that the transition matrix and class-specific choice models remained unbiased.

**Table 1: Monte Carlo Simulation Results**

| Variable | True Value | N = 5000 & T = 10 | N= 5000 & T = 5 |
|---|---|---|---|
| Initialization Probability (class 1) | 0.40 | 0.39 | 0.58 |
| Initialization Probability (class 2) | 0.60 | 0.61 | 0.42 |
| Transition Probability (class 1 \| class 1) | 0.80 | 0.78 | 0.79 |
| Transition Probability (class 2 \| class 1) | 0.20 | 0.22 | 0.21 |
| Transition Probability (class 1 \| class 2) | 0.30 | 0.29 | 0.29 |
| Transition Probability (class 2 \| class 2) | 0.70 | 0.71 | 0.71 |
| Probability(outcome 1 \| class 1) | 0.50 | 0.49 | 0.49 |
| Probability(outcome 2 \| class 1) | 0.50 | 0.51 | 0.51 |
| Probability(outcome 1 \| class 2) | 0.70 | 0.71 | 0.72 |
| Probability(outcome 2 \| class 2) | 0.30 | 0.29 | 0.28 |

## 6. Dataset

In February of 2007, Santiago, Chile introduced Transantiago, a complete redesign of the public transit system in the city. Before the introduction of Transantiago, public transport in Santiago comprised a privately operated and uncoordinated system of buses and shared taxis, and the publicly run underground Metro lines. The old bus system was characterized by a large and inefficient fleet of 8,000 buses operating 380 lines, competition among buses on streets to gain passengers, higher than required frequencies along the busiest corridors and inadequate service along the less travelled ones, low quality vehicles, high accident rates, rude drivers, high levels of air and noise pollution, fractured ownership, and many empty buses circulating during off peak hours (Yáñez et al., 2010). The Metro system, though considerably safer, faster and more reliable than the bus system, only accounted for 8% of the city's trips under the old system, due largely to sparser network coverage and the high cost of transfers between buses and the Metro.

With the aim of addressing these problems and stemming the decline in the public transportation system, the city assembled a team of Chilean specialists and consultants in 2005 to come up with a design for Transantiago (Fernández et al., 2008). Under the new system, the metropolitan region in and around Santiago was divided into ten zones and operations were taken over by a group of ten new companies. Bus routes were consolidated into a hierarchical system of trunk and feeder routes. The feeder routes connected each of these zones to the Metro lines, which served as the backbone of the new system. The trunk routes complemented the Metro lines by connecting different zones of the city. Benefits envisaged under Transantiago included the elimination of route redundancies, increased safety through the introduction of new low-floor buses, approximately half of them articulated, an integrated fare collection system through the means of a contactless smart card, lower travel times, a smaller fleet size, and reduced levels of air and noise pollution.

Though the system succeeded in achieving many of these goals, as a result of poor implementation it inadvertently created several new problems. First, the system was introduced in a 'big-bang' fashion with no pilot studies or public information campaigns leading up to the change. As a consequence, the first few weeks following the change resulted in great chaos and confusion among users of the city's public transportation system. Second, the system was designed under the assumption that by the time of its introduction, certain critical bus-only lanes would have been constructed and all buses in the public transit fleet would have been fitted with on-board GPS tracking systems. Neither of these goals was achieved in time, and as a consequence buses ran well below design speeds, introducing significant unreliability into the system. Third, most new bus routes were confined to run along major arterials, increasing the access and egress distances to bus stops, particularly in the suburban corners of the city. And finally, given the hierarchical nature of the new bus system, most bus routes were limited to run within the boundary of a single zone, increasing the number of transfers for trips that required traversing multiple zones. These four factors combined drove a number of passengers to alternative modes of travel, most notably the Metro, which, unlike the bus system, ran at least as reliably as before, resulting in extreme overcrowding on Metro trains, with average occupancy levels during peak hours on certain routes of 5-6 passengers per square meter. As one can imagine, Transantiago generated considerable ill will among city residents, some of which has persisted to this day.

The dataset for the study comes from the Santiago Panel, comprising four one-week waves of pseudo travel-diary data collected over a span of twenty-two months that extends both before and after the introduction of Transantiago. The first wave was conducted in December 2006, three months before Transantiago was introduced, and the next three waves were implemented in May 2007, December 2007 and October 2008, respectively. Survey respondents were drawn from full-time employees working at one of six campuses of Pontificia Universidad Católica de Chile spread across Santiago. Each wave of data collection had an observation period of one week, and survey respondents were asked to report the travel mode(s) that they used for their morning commute to work each day during that week. Therefore, each wave contains up to five observations per individual (corresponding to the five-day working week). Though this limits the number of destinations to just these six campuses, the panel was fortunate in that the distribution of origins was well spread across the city. In all, the Panel interviewed 303 individuals during the first wave, 286 individuals during the second wave, 279 individuals during the third wave, and 258 individuals during the final wave. Considering that the four waves were spread across nearly two years, the Panel has a comparatively low attrition rate. Each of the respondents was asked questions regarding their socioeconomic characteristics; attributes of their morning trip to work; additional activities before, during and after work and their influence, if any, on the respondent's choice of travel mode; subjective perceptions about the performance of the new system (collected only during the second and third waves); and their level of agreement with attitudinal statements about different aspects of the transportation system, such as safety, reliability and accessibility (collected only during the fourth wave). For more details about the dataset, the reader is referred to Yáñez et al. (2010).

The dataset offers a unique opportunity to investigate the effects of systemic changes in the transportation network on the evolution and persistence of individual preferences. For the purpose of our analysis, we will be restricting our attention to 220 respondents, each of whom has at least one recorded observations in each of the four waves that constitute the Panel. We aggregate the modal alternatives into seven travel modes: auto, metro, bus, walk, bike, auto/metro (for individuals that drive to the metro station, and take the metro from there), and bus/metro (for individuals that take the bus to the metro station, and the metro from there).

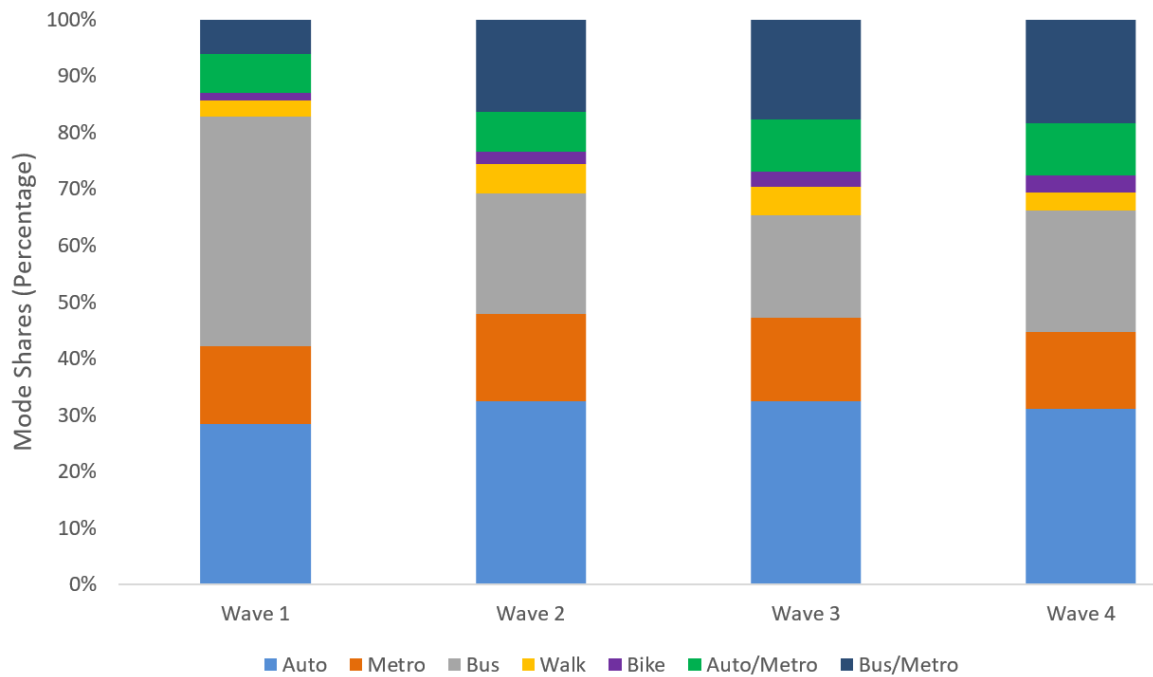

**Figure 3: Mode Shares across All Waves**

Figure 3 plots mode shares across the four waves for all 220 individuals. It is evident that there was a big reduction in choosing the bus system as a mode of transport for work trips after wave one (post introduction of Transantiago). Bus mode shares declined from 40.6% during wave one to 18.2% during wave three, before marginally rebounding to 21.6% during wave four. Mode shares for auto/metro and bus/metro increased dramatically after the introduction of Transantiago. The major shifts in the mode choices occurred between waves one and two, as one would expect. Shifts tend to stabilize over time as people get more adjusted with their new work trips mode choice habits.

The reader should note that a plot like Figure 3 could also have been plotted using repeated cross-sectional data. Longitudinal data allows us to analyze where these changes in mode shares are coming from. Figure 4 plots the number of trips where individuals switched travel modes between any two subsequent waves of the Panel. The scale of the vertical axes for each of the three plots is the same, to make the comparison easier. As one would expect, the majority of the shift occurs from wave 1 to wave 2, immediately in the wake of the introduction of Transantiago, and most of it away from "bus" and towards

"bus/metro". However, as the system stabilizes over time, so does the behavior of its users, with significantly less movement across travel modes between waves 2 and 3 and waves 3 and 4. Given the nature of the differences between the old and the new system, this is not surprising. The more interesting question is: does the shift in observable travel mode choice behavior indicate a corresponding shift in latent travel mode preferences? And does this latter shift, if any, persist beyond the first wave? We address these related questions using the HMM framework in the next section.

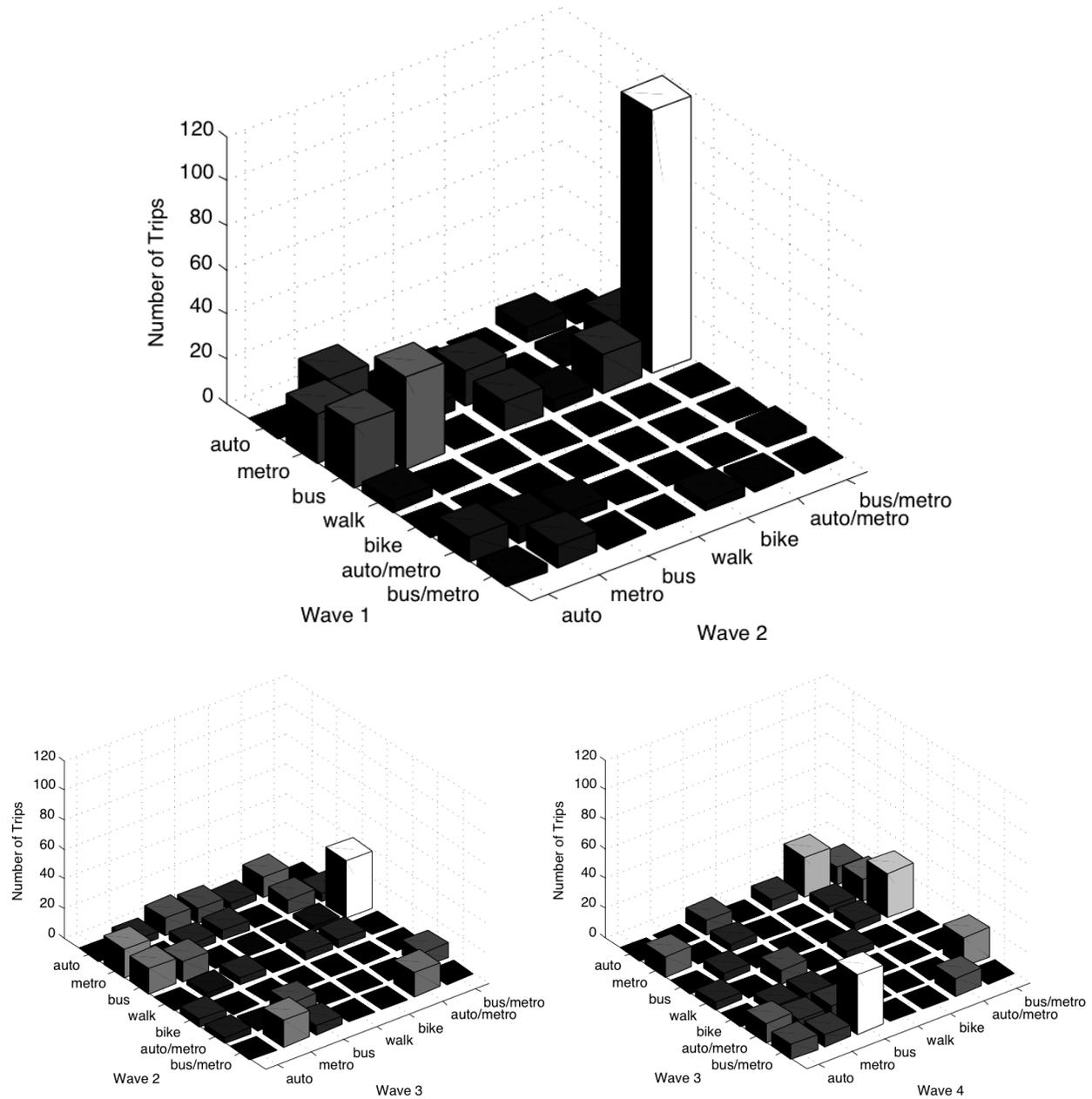

**Figure 4: Shifts across Travel Modes between Subsequent Waves of the Panel**

El Zarwi, Vij, and Walker                                                                                                      19Finally, when estimating HMMs, time periods should ideally be evenly spaced. In our case, the time periods are denoted by each wave of data collection. As mentioned before, the interval between waves is not evenly spaced: five months between waves 1 and 2, seven months between waves 2 and 3, and ten months between waves 3 and 4. Given that the main objective of this paper is to develop a framework for modeling and forecasting the evolution of preferences over time, we need to assume that the transition model parameters are stable. With unevenly spaced time periods, we need to account for these differences in the transition model specification, explicitly or implicitly. One could estimate a heterogeneous HMM, where the transition model parameters are specified as an explicit function of the time interval between waves. Alternatively, one could estimate a homogenous hidden Markov model with time independent transition model parameters, such that the parameters are implicitly averaged over the different time intervals. For the sake of simplicity, we adopted the implicit approach. Therefore, when using the model to forecast changes in preferences and behaviors beyond wave four, the time intervals between future waves will be taken as the average time interval between successive waves for the first four waves.

## 7. Estimation Results and Discussion

The following section presents results from the hidden Markov model. Our proposed dynamic discrete choice framework models the evolution of preferences over time in response to changes in socio-demographic variables and the level-of-service of the transportation network. Determining the final model specification was based on varying the utility specification for all sub-models: initialization model, transition model and class-specific choice model. The method for identifying the number of distinct preference states i.e. modality styles that exist in the sample population, is iterative. The models were built incrementally: we first estimated a model with two modality styles, using that as a starting point for the model with three modality styles, and so on. The final number of modality styles in our sample was determined based on a comparison across measures of statistical fit, such as the rho-bar-squared ($\bar{\rho}^2$), Akaike Information Criterion (AIC) and Bayesian Information Criterion (BIC), and behavioral interpretation.

We first estimated the models by varying the number of modality styles excluding the effect of the level-of-service of the transportation network on the modality style transition model, as captured through the construct of consumer surplus (i.e. the average expected maximum utility, or inclusive value). We made use of the power of the EM algorithm in estimating model parameters while saving on computation time. The EM algorithm provides a statistically robust approach for model estimation by taking advantage of the conditional independence structure of the model framework. We estimated models with two, three and four modality styles. Table 2 enumerates the statistical measures of fit for each of these models. While the AIC and the BIC decrease as the number of classes increases, the rho-bar-squared value is highest for the three class model. However, a joint comparison across both statistical measures of fit and behavioral interpretation led us to select the four class model as the preferred specification. The four class model was subsequently reestimated, adding the measure of consumer surplus from the class-specific mode choice models to the transition model.

**Table 2: Measures of Model Fit**

| Number of Modality Styles | Log-Likelihood | $\bar{\rho}^2$ | AIC | BIC |
|---|---|---|---|---|
| Two | -2365 | 0.514 | 4798 | 5015 |
| Three | -1599 | 0.636 | 3328 | 3743 |
| Four | -1287 | 0.601 | 2740 | 3270 |

Tables 3, 4 and 5 present detailed parameter estimates (and t-statistics) of the class-specific travel model choice model, initialization model and transition model, respectively, for the final specification. The four classes, or modality styles, differ from each other in terms of the travel modes that they consider, their sensitivity to the level-of-service of the transportation system, and their socio-demographic composition over time. The tabulated model results are behaviorally consistent, i.e. parameter estimates across all sub-models, and in particular the class-specific travel mode choice model, have the expected sign and are statistically significant. Over subsequent paragraphs, we summarize key characteristics of each of the classes. To underscore behavioral differences between classes, a sample enumeration is carried out across the four waves, and the results are incorporated in our description of the classes.

El Zarwi, Vij, and Walker    21El Zarwi, Vij, and Walker    21

**Table 3: Class-specific Travel Mode Choice Model Results**

| Variable | Class 1 Drivers | Class 2 Bus Users (Transit) | Class 3 Bus-Metro Users (Transit) | Class 4 Auto-Metro Users (Transit) |
|---|---|---|---|---|
| Alternative Specific Constant | | | | |
| Auto | 0.000 (-) | - | - | 0.000 (-) |
| Metro | -3.925 (-10.134) | - | 0.000 (-) | 2.293 (208.455) |
| Bus | -4.259 (-13.437) | 0.000 (-) | -7.644 (-19.739) | - |
| Walk | 1.935 (6.158) | - | - | - |
| Bike | -0.710 (-3.214) | - | - | - |
| Auto-Metro | -3.440 (-11.576) | - | - | 4.441 (753.179) |
| Bus-Metro | -3.618 (-9.750) | - | 2.208 (6.965) | - |
| Travel Time (mins) | -0.028 (-2.968) | -0.042* (-0.275) | -0.091* (-0.290) | -0.069 (-3.043) |
| Walk Time (mins) | -0.041 (-3.761) | -0.002* (-0.019) | -0.127* (-0.574) | -0.103* (-0.073) |
| Travel Cost (CLP) | -0.006* (-1.072) | -0.061* (-0.280) | -0.102* (-0.344) | -0.080* (-0.074) |
| Waiting Time (mins) | -0.024* (-1.065) | -0.038* (-0.042) | -0.293* (-0.790) | -0.053* (-0.940) |
| Number of Transfers | - | -2.633 (-13.894) | -1.136 (-118.488) | - |

- Not applicable; * Insignificant at the 5% level

**Table 4: Initialization Model Results**

| Variable | Class 1 Drivers | Class 2 Bus Users (Transit) | Class 3 Bus-Metro Users (Transit) | Class 4 Auto-Metro Users (Transit) |
|---|---|---|---|---|
| **Initialization Model (Wave 1)** | | | | |
| Alternative Specific Constant | 0.000 (-) | 2.993 (5.951) | -0.073* (-0.160) | 0.139* (0.229) |
| Household Income (100,000s CLP) | 0.000 (-) | -0.510 (-4.621) | -0.060* (-1.313) | -0.190 (-2.008) |
| Male | 0.000 (-) | 0.223* (0.521) | 0.635* (1.176) | 0.519* (0.821) |
| Number of Vehicles | 0.000 (-) | -0.992 (-3.159) | -0.739 (-1.979) | -0.295* (-0.736) |

- Not applicable; * Insignificant at the 5% level

**Table 5: Transition Model Results**

| Variable | Class 1 Drivers | Class 2 Bus Users (Transit) | Class 3 Bus-Metro Users (Transit) | Class 4 Auto-Metro Users (Transit) |
|---|---|---|---|---|
| **Transition Model (Given Class 1 in Wave t- 1)** | | | | |
| Alternative Specific Constant | 0.000 (-) | 0.900 (28.746) | 1.351 (12.388) | -2.175 (-44.022) |
| Household Income (100,000s CLP) | 0.000 (-) | -0.170* (-0.002) | -0.610* (-0.010) | -0.080* (-0.001) |
| Male | 0.000 (-) | 0.671 (21.067) | -1.178 (-21.928) | 0.359 (21.710) |
| Number of Vehicles | 0.000 (-) | -0.385 (-5.752) | -0.416* (-0.262) | -0.365* (-0.221) |
| Consumer Surplus (utils) | 0.594* (0.303) | 1.000 (-) | 0.264 (43.803) | - |
| **Transition Model (Given Class 2 in Wave t- 1)** | | | | |
| Alternative Specific Constant | 0.000 (-) | 4.833 (7.617) | 2.060 (1637.711) | 1.223 (382.495) |
| Household Income (100,000s CLP) | 0.000 (-) | -0.680 (-10.155) | -0.310* (-0.003) | -0.500* (-0.005) |
| Male | 0.000 (-) | 1.831 (3.014) | 1.056* (1.477) | 0.999* (1.506) |
| Number of Vehicles | 0.000 (-) | 0.595* (1.135) | -0.130* (-0.100) | -0.378* (-0.248) |
| Consumer Surplus (utils) | 0.330 (110.466) | 0.500 (256.703) | 0.155* (0.114) | 0.317* (0.253) |
| **Transition Model (Given Class 3 in Wave t- 1)** | | | | |
| Alternative Specific Constant | 0.000 (-) | 2.480* (1.371) | 0.936 (414.323) | 1.391 (626.478) |
| Household Income (100,000s CLP) | 0.000 (-) | -1.150 (-2.811) | -0.090* (-0.001) | -0.930* (-0.012) |
| Male | 0.000 (-) | 0.635* (0.560) | 1.801 (3.300) | -0.641* (-1.087) |
| Number of Vehicles | 0.000 (-) | -1.506* (-1.495) | -1.143* (-0.608) | 0.184* (0.143) |
| Consumer Surplus (utils) | 1.709* (1.688) | 0.140* (0.098) | 0.097* (0.108) | 0.364* (0.560) |
| **Transition Model (Given Class 4 in Wave t- 1)** | | | | |
| Alternative Specific Constant | 0.000 (-) | 1.064* (0.504) | 0.636 (123.621) | 0.968 (1003.253) |
| Household Income (100,000s CLP) | 0.000 (-) | -0.370* (-0.712) | -0.060* (-0.001) | 0.050* (0.001) |
| Male | 0.000 (-) | 1.84* (1.07) | 1.03* (1.10) | 0.04* (0.06) |
| Number of Vehicles | 0.000 (-) | -1.805* (-1.905) | -0.163* (-0.058) | 0.142* (0.061) |
| Consumer Surplus (utils) | 0.088* (0.513) | 0.094* (0.501) | 0.083* (0.487) | - |

- Not applicable; * Insignificant at the 5% level

**Class 1 (drivers):** This class constitutes 36% of the sample population during wave 1, and the share of the class slowly but steadily increases to 40% by wave 4. Individuals belonging to this class consider all available modes of transport, but 70% of all trips are made by auto. Value of time varies between 0.36$/hr for waiting time and 0.62$/hr for walking time, and the class is completely insensitive to public transport transfers. High-income men with cars are most likely to belong to this class.

**Class 2 (bus users):** This class constitutes 39% of the sample population during wave 1, and the share of the class steadily decreases to 20% by wave 4. Individuals belonging to this class deterministically choose bus for all their trips. Value of time is low, at approximately 0.07$/hour across different travel time components. Note that even though the class-specific choice model is deterministic, parameters denoting sensitivities to travel times and costs can still be estimated indirectly through the transition model through the construct of consumer surplus. High-income individuals with cars are most likely to belong to this class initially, but they are also most likely to leave this class after the introduction of Transantiago.

**Class 3 (bus-metro users):** This class constitutes 14% of the sample population during wave 1, and the share of the class steadily increases to 24% by wave 4. Individuals belonging to this class consider the metro, bus and bus-metro alternatives. Value of time varies between 0.09$/hr for in-vehicle time and 0.26$/hr for waiting time. Each public transport transfer is equivalent to 12 minutes of in-vehicle time. Low-income women without access to cars are most likely to belong to this class.

**Class 4 (auto-metro users):** This class constitutes 11% of the sample population during wave 1, and the share of the class increases marginally to 16% by wave 4. Individuals belonging to this class consider the auto, metro and auto-metro alternatives. Value of time varies between 0.06$/hr for waiting time and 0.12$/hr for walking time, and the class is completely insensitive to public transport transfers. While low-income individuals without access to cars are most likely to belong to this class initially, over time, more high-income individuals with access to cars migrate to this class.

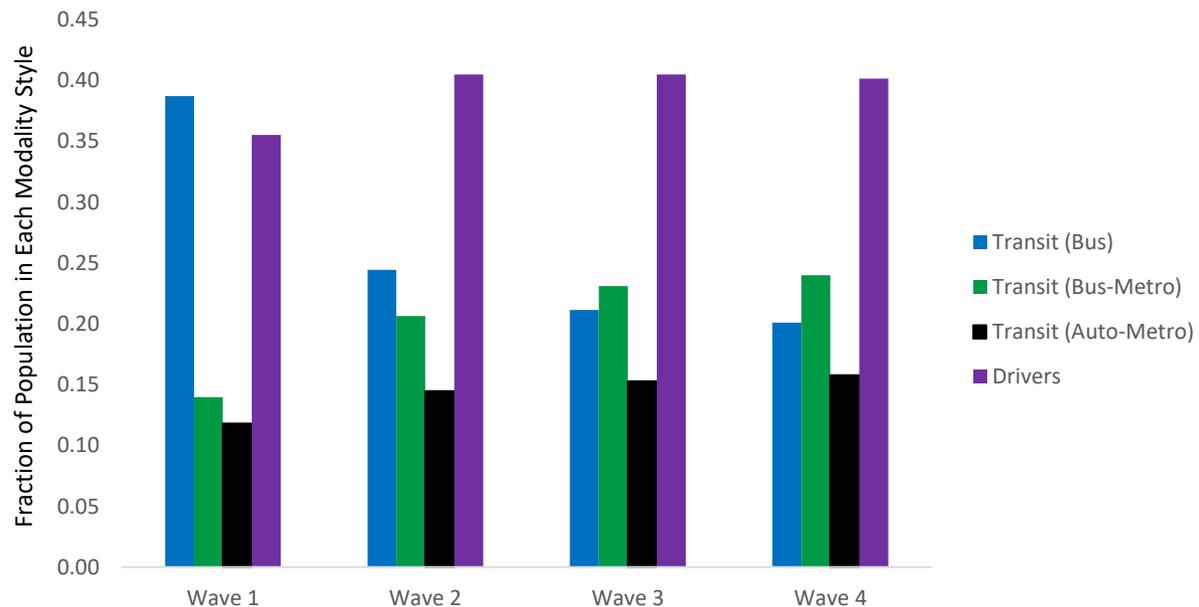

**Figure 5: Estimated Share of Individuals in Each Modality Style across Waves**

Now that we have estimated our hidden Markov model, we want to explore the power of this model in terms of explaining the evolution of preferences, or modality styles, in response to the introduction of Transantiago. The population distribution of individuals across the four classes for each of the waves, as determined by sample enumeration, is displayed in Figure 5. It is evident that a shock to the transportation network along the lines of Transantiago did force people to reconsider their modes for travel. The market share of drivers, bus-metro, and auto-metro users has increased after the introduction of Transantiago, while the market share for bus users has drastically decreased. These results are aligned with findings from Section 6 regarding mode share percentages of the different modes across the four waves. We can see that major reductions and increases in shares of modality styles occurred right after Transantiago revolutionized the public transit system. These population changes stabilize over time. It is also evident from the figure that population preferences have in fact changed over time, and in particular after the introduction of Transantiago.

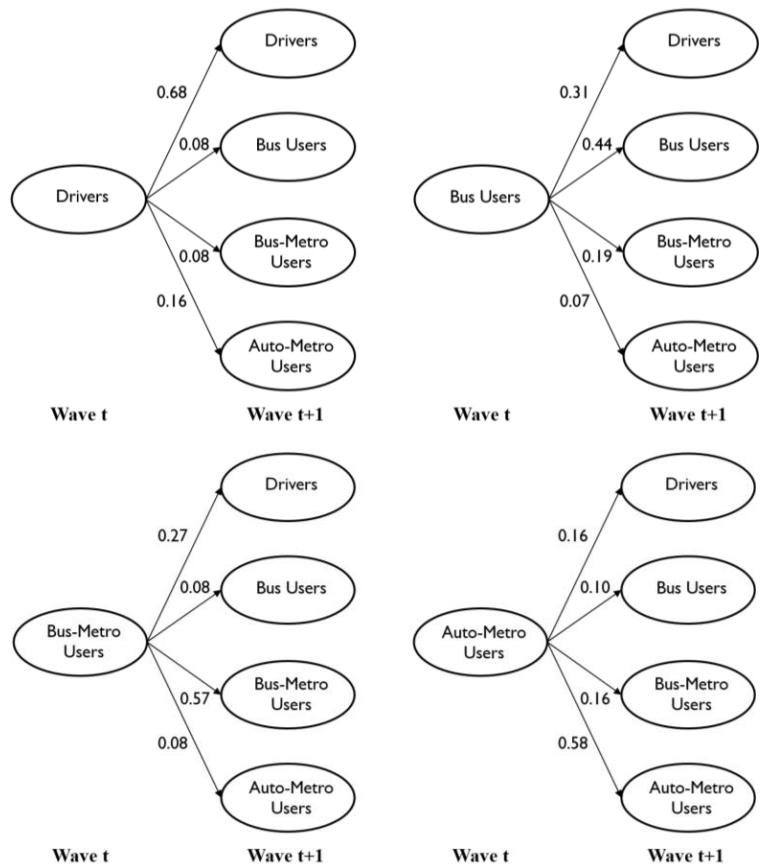

**Figure 6: Estimated Average Transition Probabilities across Modality Styles over Time**

However, the stability in preferences at the population level belies the instability at the individual level. Figure 6 illustrates the average transition probabilities between different modality styles across successive time periods. Note that while Figure 5 could have been reproduced using a static framework, such as an LCCM, with repeated cross-sectional data (see, for example, Vij et al., 2017), Figure 6 could only be produced using a dynamic framework with longitudinal data, such as the HMM proposed here. Interestingly, transition probabilities were not found to differ substantially across time periods, and for this reason, we present average values over all time periods. There are two key trends to note here. First, construct of habit formation is implicitly captured in the relative magnitude of the transition probabilities. In general, decision-makers are more inclined to remain in the same modality style over time than switch to a different modality style. For three of the four modality styles, the probability of staying in that modality style over successive time periods is found to be greater than half. And second, there is considerable instability in travel mode preferences, despite the relative stability at the population level and the strong influence of habit at the individual level. For example, roughly 30% of bus users and bus-metro users become drivers each time period. Part of this transition could be explained by the introduction of Transantiago, which did make use of the public transport system in a more onerous manner. However, the trend persists beyond wave 2, several months after the introduction of Transantiago, indicating a more general and ongoing shift in preferences towards the car over time, triggered possibly in part by the introduction of Transantiago.

## 8. Policy Analysis

Practitioners and policy analysts are often interested in understanding and predicting broad population trends in travel and activity behavior. Does failure to account for preference dependencies over time impact population estimates? Or is it reasonable to ignore such dependencies when undertaking population-level analysis? We address these questions by comparing aggregate forecasts from the HMM with static frameworks that do not account for preference dependencies over time. The forecasting horizon is limited to three waves post the fourth wave (i.e. waves five, six and seven). We simulate the following two policy scenarios:

1- Increasing household income by 10% at waves five, six and seven respectively.
2- Reducing travel time by 15% for the bus and bus to metro alternatives. This could be brought about by a new transportation policy, dedicated bus lanes for example. This particular shock to the transportation network is assumed to take place between waves four and five.

We compare forecasts from the HMM with latent class choice models (LCCMs). To ensure that the LCCMs and the HMM are as similar as possible, and any potential differences in forecasts cannot be attributed to differences in either observed data or model specification, we use the following procedure. Since an LCCM would typically be estimated using a single cross-section, we estimate two separate LCCMs using data from the first and last wave respectively. Each of the LCCMs comprises four modality styles (preference states), same as our HMM. We constrain the class-specific choice model for each LCCM to be the same as that of the HMM. We only estimate the class membership model parameters, where we formulate class membership as a function of socio-demographic variables, namely income, gender and level of car ownership, and the consumer surplus offered by each class. These are the same variables that are included in the specification of the transition model for the HMM.

Figure 7 plots the change in modality styles across waves five, six and seven for the first policy scenario, as predicted by the HMM and the two LCCMs, and Figure 8 plots the corresponding travel mode shares for the same. As is evident from the figures, even at the population level, there are considerable differences between forecasts from the three models. In general, the LCCM estimated using wave 4 data more closely tracks forecasts from the HMM. Relative to the HMM, the LCCM estimated using wave 4 data under predicts the share of drivers and auto-metro users, and over predicts the share of bus-metro users, whereas the LCCM estimated using wave 1 data under predicts the share of bus-metro users, and over predicts the share of drivers and bus users. These differences translate into similar inconsistencies in travel mode shares. For example, travel mode shares for the pure public transport modes, i.e. bus, metro and bus to metro, during wave 5 are predicted to be 49% by the HMM, 45% by the LCCM estimated using wave 1 data, and 54% by the LCCM estimated using wave 4 data.

Figures 9 and 10 plot corresponding forecasts for the second policy scenario, as predicted by the HMM and the two LCCMs. Note that the travel mode shares are the same across all three waves, since the change in the transportation system precedes wave 5. Therefore, we show them as a single plot. In terms of modality styles, there are considerable differences between forecasts from the three models, though forecasts from the LCCM estimated using wave 4 data are in closer agreement with those from the HMM. Interestingly, changes in in-vehicle travel times between waves 4 and 5 do not have a significant impact on the likelihood of belonging to a particular modality style over subsequent waves, as predicted by each of the three models, and the population distribution remains largely unchanged across waves 5, 6 and 7.

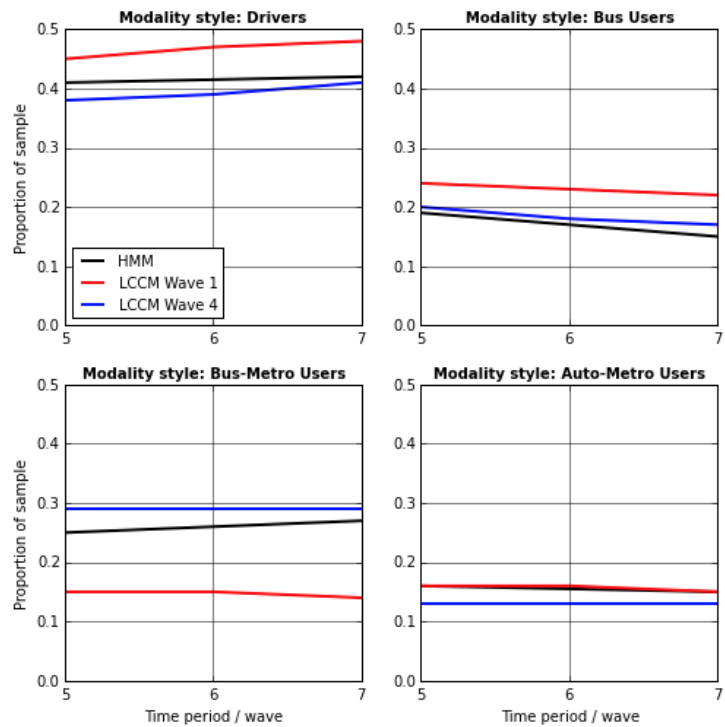

**Figure 7: Share of Individuals in Each Modality Style for Policy Scenario 1, as Predicted by the HMM, the LCCM Estimated Using Wave 1 Data, and the LCCM Estimated Using Wave 4 Data**

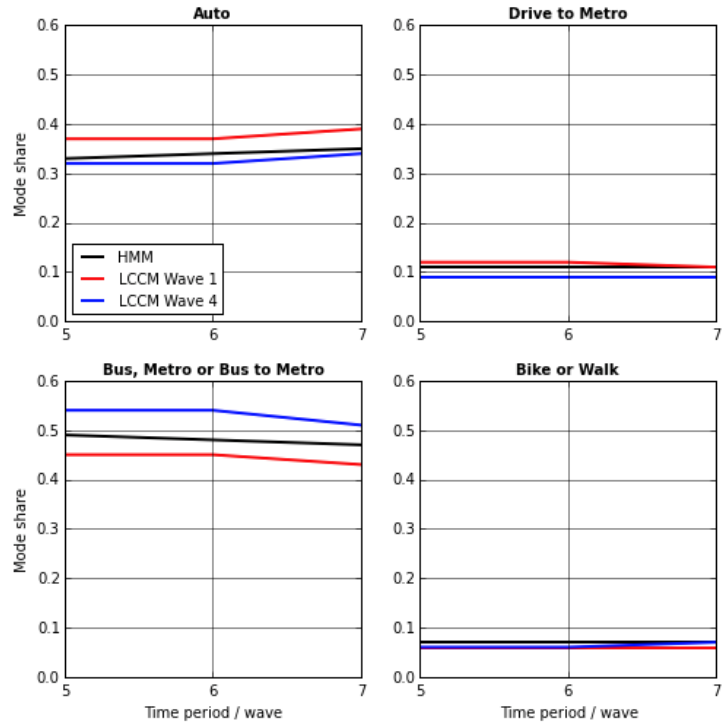

**Figure 8: Mode Shares for Policy Scenario 1, as Predicted by the HMM, the LCCM Estimated Using Wave 1 Data, and the LCCM Estimated Using Wave 4 Data**

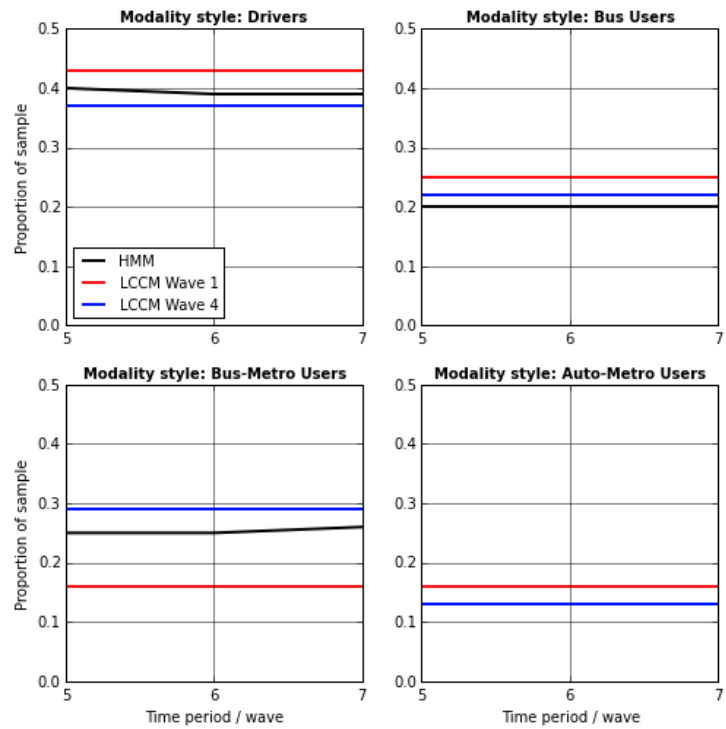

**Figure 9: Share of Individuals in Each Modality Style for Policy Scenario 2, as Predicted by the HMM, the LCCM Estimated Using Wave 1 Data, and the LCCM Estimated Using Wave 4 Data**

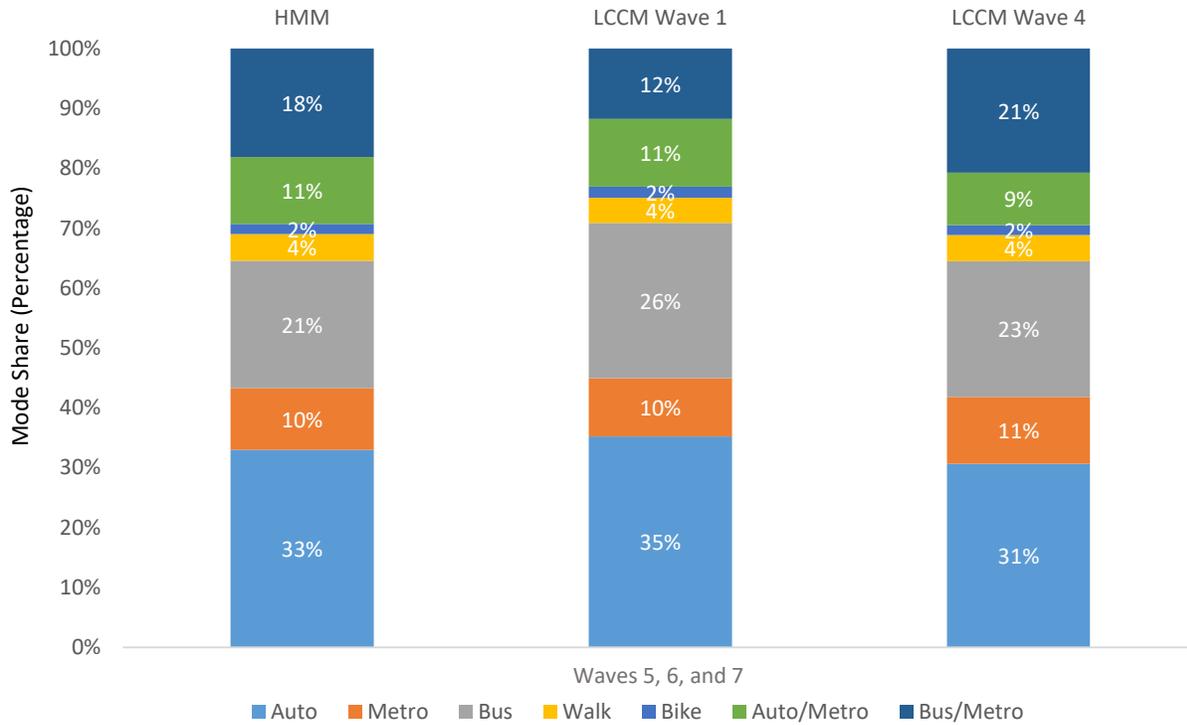

**Figure 10: Mode Shares for Policy Scenario 2, as Predicted by the HMM, the LCCM Estimated Using Wave 1 Data, and the LCCM Estimated Using Wave 4 Data**

It is important to note that across both scenarios, the predicted share of drivers and bus users has been strictly higher via the LCCM estimated using wave 1 data, compared to the other two models, while the share of bus-metro users has been significantly lower. The reason behind that is the fact that observations pertinent to wave one (before the introduction of Transantiago) constituted a sample of the population that preferred taking the bus or driving to work. Moreover, the market share for the metro alternative was significantly lower during wave one. In addition to that, forecasts from the HMM and LCCM estimated using wave 4 data seem to be more consistent with each other in terms of the evolutionary trends of preferences. However, the share of individuals in the four preference states tends to be different.

In terms of aggregate mode shares, differences across the three models are equally sizeable. For example, travel mode shares for bus are predicted to be 21% by the HMM, 26% by the LCCM estimated using wave 1 data, and 23% by the LCCM estimated using wave 4 data. And similarly, travel mode shares for bus to metro are predicted to be 18% by the HMM, 12% by the LCCM estimated using wave 1 data, and 21% by the LCCM estimated using wave 4 data. As we argued before, relative to the HMM, the LCCM estimated using wave 1 data over predicts mode shares for bus and under predicts mode shares for bus to metro, and the LCCM estimated using wave 4 data over predicts mode shares for bus to metro. These differences are not unexpected. Bus use was at its greatest during the first wave of observation. And subsequent structural changes in the public transportation system, initiated by Transantiago, increased the popularity of bus to metro over the following waves. On one hand, the LCCM estimated using wave 1 data is unable to predict the full extent of changes in behavior in response to these changes in the transportation system. On the other, the LCCM estimated using wave 4 data overstates these changes in behavior, as it does not account for habit formation from preferences and behaviors that precede Transantiago.

## 9. Conclusions

The objective of this study was to develop a methodological framework that can model and forecast the evolution of individual preferences and behaviors over time. Traditionally, discrete choice models have formulated preferences as a function of demographic and situational variables, psychological, sociological and biological constructs, and available alternatives and their attributes. However, the impact of past experiences on present preferences has usually been overlooked.

We developed a hidden Markov model (HMM) of travel mode choice behavior. The hidden states denote travel mode preferences, or modality styles, that differ from one another in terms of the travel modes considered when deciding how to travel, and the relative sensitivity to different level-of-service attributes of the transportation system. The evolutionary path is assumed to be a first-order Markov process, such that an individual's modality style during a particular time period depends only on their modality style in the previous time period. Transitions between modality styles over time are assumed additionally to depend on changes in demographic variables and the transportation infrastructure (available travel modes and their attributes). Conditional on the modality styles that an individual has during a particular time period, the individual is assumed to choose that travel mode which offers the greatest utility.

The model framework was empirically evaluated using data from the Santiago Panel. The dataset comprises four waves of one-week pseudo travel diaries each. The first wave was conducted before the introduction of Transantiago, a complete redesign of the public transit system in Santiago, Chile, and the next three waves were conducted after. The dataset offered a unique opportunity to study the impact of a shock to the transportation network on the stability of travel mode preferences over time. The model identified four modality styles in the sample population: drivers, bus users, bus-metro users and auto-metro users. At the population level, the proportion of drivers, auto-metro users, and bus-metro users has increased after the introduction of Transantiago, and the proportion of bus users has drastically decreased. The biggest shift happens between the first and second wave, the same period when Transantiago is introduced. The population distribution is more or less stable across the latter three waves. However, at the individual level, we observe two interesting phenomena. First, habit formation is found to impact transition probabilities across all modality styles. Individuals are more likely to stay in the same modality style over successive time periods than transition to a different modality style. And second, despite both the stability in preferences at the population level and the influence of habit formation at the individual level, nearly 40% of the sample population is found to change modality styles between any two successive waves, reflecting great instability in individual preferences, much after the introduction of Transantiago. These findings hold implications for aggregate forecasts. We simulated two policy scenarios using the HMM, and two latent class choice model (LCCM) framework with comparable specifications, estimated using two separate cross-sections of the Santiago Panel. Relative to the HMM, the first LCCM, estimated using data from before the introduction of Transantiago, under predicts changes in travel mode shares, due to its inability to observe the potential impact of a transformative change such as Transantiago. Relative to the HMM, the second LCCM, estimated using data from after the introduction of Transantiago, over predicts changes in travel mode shares, due to its inability to account for habit formation of preferences and behaviors from before the introduction of Transantiago.

There are two key directions in which future research can build on findings from this study. First, the methodological framework developed here captures preference dependencies across time for the same individual, explicitly accounting for the effect of habit formation on travel behavior. The framework offers the potential to improve the accuracy of the long-range forecasts made from large-scale urban travel demand models. Future research should explore ways in which existing travel demand modeling paradigms can adopt dynamic representations of behavior that capture temporal trends in preferences and behaviors. And second, the framework developed here provides a quantitative basis for modeling and forecasting structural shifts in preferences that are bound to occur in this era of transformative mobility. We observed great flux in individual commute travel mode preferences over time, triggered at least in part by a major redesign of the public transportation system. It would be interesting to see how these findings compare with corresponding changes in preferences across other dimensions of travel behavior, such as car ownership and residential location, and in response to other changes in transport policy, infrastructure and services, such as the introduction of congestion charge schemes, the diffusion of alternative-fuel vehicles and the emergence of shared mobility services.

## Acknowledgements

We would like to thank Juan de Dios Ortúzar for providing us access to the Santiago panel, which was used to estimate our dynamic choice model; and Tony Marley, for very thoughtful and detailed feedback on an earlier draft of the paper.

El Zarwi, Vij, and Walker                                                                                32